\documentclass[
showpacs,
floatfix,
aps,
prl,
amsmath,
twocolumn,
superscriptaddress,
tightenlines
groupaddress,
]{revtex4}

\usepackage{graphicx}
\usepackage{dcolumn}
\usepackage{amsmath}
\usepackage{amsfonts}
\usepackage{bm}
\usepackage{times}
\usepackage[varg]{txfonts}

\def\tautr{\tau_{\mbox{\scriptsize tr}}}
\def\tauq{\tau_{\mbox{\scriptsize q}}}
\def\tauin{\tau_{\mbox{\scriptsize in}}}
\def\tauim{\tau_{\mbox{\scriptsize {q}}}^{\mbox{\scriptsize {im}}}}

\def\tauee{\tau_{\mbox{\scriptsize {q}}}^{\mbox{\scriptsize {ee}}}}
\def\taubar{\bar\tau}

\def\eac{\epsilon_{\mbox{{\scriptsize ac}}}}

\def\pac{\phi_{\mbox{\scriptsize ac}}}

\def\oc{\omega_{\mbox{\scriptsize {c}}}}

\def\pc{{\bar{\cal P}}}
\def\ec{{\cal E}_\omega}

\def\ain{A_{\mbox{\scriptsize in}}}
\def\adis{A_{\mbox{\scriptsize dis}}}

\bibpunct{[}{]}{,\!}{n}{,}{,} 

\begin{document}
\title{
Temperature Dependence of Microwave Photoresistance in 2D Electron Systems
}
\author{A.\,T. Hatke}
\affiliation{School of Physics and Astronomy, University of Minnesota, Minneapolis, Minnesota 55455, USA} 
\author{M.\,A. Zudov}
\email[Corresponding author: ]{zudov@physics.umn.edu}
\affiliation{School of Physics and Astronomy, University of Minnesota, Minneapolis, Minnesota 55455, USA} 
\author{L.\,N. Pfeiffer}
\affiliation{Bell Labs, Alcatel-Lucent, Murray Hill, New Jersey 07974, USA}
\author{K.\,W. West}
\affiliation{Bell Labs, Alcatel-Lucent, Murray Hill, New Jersey 07974, USA}

\begin{abstract}
We report on the temperature dependence of microwave-induced resistance oscillations in high-mobility two-dimensional electron systems.
We find that the oscillation amplitude decays exponentially with increasing temperature, as $\exp(-\alpha T^2)$, where $\alpha$ scales with the inverse magnetic field.
This observation indicates that the temperature dependence originates {\em primarily} from the modification of the single particle lifetime, which we attribute to electron-electron interaction effects.
\end{abstract}
\vspace{-0.4in}
\maketitle

Over the past few years it was realized that magnetoresistance oscillations, other than Shubnikov-de Haas oscillations \cite{shubnikov:1930}, can appear in high mobility two-dimensional electron systems (2DES) when subject to microwaves \cite{miro:exp}, dc electric fields \cite{yang:2002a}, or elevated temperatures \cite{zudov:2001b}.
Most attention has been paid to the microwave-induced resistance oscillations (MIRO), in part, due to their ability to evolve into zero-resistance states \citep{mani:2002,zudov:2003,willett:2004,zrs:other}.
Very recently, it was shown that a dc electric field can induce likely analogous states with zero-differential resistance \citep{bykov:zhang}.

Despite remarkable theoretical progress towards understanding of MIRO, several important experimental findings remain unexplained. 
Among these are the immunity to the sense of circular polarization of the microwave radiation \citep{smet:2005} and the response to an in-plane magnetic field \citep{mani:yang}.
Another unsettled issue is the temperature dependence which, for the most part \citep{studenikin:2007}, was not revisited since early reports focusing on the apparently activated behavior of the zero-resistance states \citep{mani:2002,zudov:2003,willett:2004}.
Nevertheless, it is well known that MIRO are best observed at $T \simeq 1$ K, and quickly disappear once the temperature reaches a few Kelvin.

MIRO originate from the inter-Landau level transitions accompanied by microwave absorption and are governed by a dimensionless parameter $\eac\equiv\omega/\oc$  ($\omega=2\pi f$ is the microwave frequency, $\oc=eB/m^*$ is the cyclotron frequency) with the maxima$^+$ and minima$^-$ found \citep{miro:phase} near $\eac^{\pm}=n \mp \pac,\,\pac \leq 1/4$ ($n \in \mathbb{Z}^+$).
Theoretically, MIRO are discussed in terms of the ``displacement'' model \citep{disp:th}, which is based on microwave-assisted impurity scattering, and the ``inelastic'' model \cite{dorozhkin:2003,dmp,dmitriev:2005}, stepping from the oscillatory electron distribution function.
The correction to the resistivity due to either ``displacement'' or ``inelastic'' mechanism can be written as \citep{dmitriev:2005}:
\begin{equation}
\delta \rho=-4\pi\rho_0\tautr^{-1}\pc\eac \taubar \delta^{2}\sin(2\pi\eac)
\label{theory}
\end{equation}
Here, $\rho_0\propto 1/\tautr$ is the Drude resistivity, $\tautr$ is the transport scattering time, $\pc$ is a dimensionless parameter proportional to the microwave power, and $\delta=\exp(-\pi\eac/\omega\tauq)$ is the Dingle factor.
For the ``displacement'' mechanism $\taubar=3\tauim$, where $\tauim$ is the long-range impurity contribution to the quantum (or single particle) lifetime $\tauq$.
For the ``inelastic'' mechanism  $\taubar=\tauin \simeq \varepsilon_F T^{-2}$, where $\varepsilon_F$ is the Fermi energy.
It is reasonable to favor the ``inelastic'' mechanism over the ``displacement'' mechanism for two reasons. 
First, it is expected to dominate the response since, usually, $\tauin \gg \tauim$ at $T\sim 1$ K.
Second, it offers plausible explanation for the MIRO temperature dependence observed in early \citep{mani:2002,zudov:2003} and more recent \citep{studenikin:2007} experiments.

In this Letter we study temperature dependence of MIRO in a high-mobility 2DES. 
We find that the temperature dependence originates primarily from the temperature-dependent quantum lifetime, $\tauq$, entering $\delta^2$.
We believe that the main source of the modification of $\tauq$ is the contribution from electron-electron scattering.
Furthermore, we find no considerable temperature dependence of the pre-factor in Eq.\,(1), indicating that the ``displacement'' mechanism remains relevant down to the lowest temperature studied.
As we will show, this can be partially accounted for by the effect of electron-phonon interactions on the electron mobility and the interplay between the two mechanisms.
However, it is important to theoretically examine the influence of the electron-electron interactions on single particle lifetime, the effects of electron-phonon scattering on transport lifetime, and the role of short-range disorder in relation to MIRO. 

While similar results were obtained from samples fabricated from different GaAs/Al$_{0.24}$Ga$_{0.76}$As quantum well wafers, all the data presented here are from the sample with density and mobility of $\simeq 2.8 \times 10^{11}$ cm$^{-2}$ and $\simeq 1.3 \times 10^7$ cm$^2$/Vs, respectively.
Measurements were performed in a $^3$He cryostat using a standard lock-in technique.
The sample was continuously illuminated by microwaves of frequency $f=81$ GHz.
The temperature was monitored by calibrated RuO$_2$ and Cernox sensors.

\begin{figure}[t]
\includegraphics{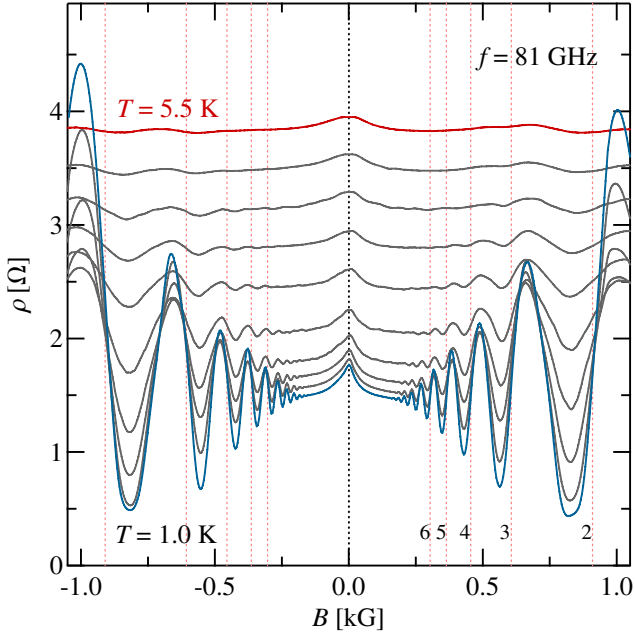}
\caption{(color online)  
Resistivity $\rho$ vs. $B$ under microwave irradiation at $T$ from 1.0 K to 5.5 K (as marked), in 0.5 K steps.
Integers mark the harmonics of the cyclotron resonance.
}
\label{fig1}
\end{figure}
In Fig.\,\ref{fig1} we present resistivity $\rho$ as a function of magnetic field $B$ acquired at different temperatures, from $1.0$ K to $5.5$ K in 0.5 K increments.
Vertical lines, marked by integers, label harmonics of the cyclotron resonance.
The low-temperature data reveal well developed MIRO extending up to the tenth order. 
With increasing $T$, the zero-field resistivity exhibits monotonic growth reflecting the crossover to the Bloch-Gr\"uneisen regime due to excitation of acoustic phonons \citep{stormer:mendez}.
Concurrently, MIRO weaken and eventually disappear at higher temperatures.
This disappearance is not due to the thermal smearing of the Fermi surface, known to govern the temperature dependence of the Shubnikov-de Haas oscillations.

We start our analysis of the temperature dependence by constructing Dingle plots and extracting the quantum lifetime $\tauq$ for different $T$. 
We limit our analysis to $\eac\gtrsim 3$ for the following reasons.
First, this ensures that we stay in the regime of the overlapped Landau levels, $\delta \ll 1$.
Second, we satisfy, for the most part, the condition, $T > \oc$, used to derive Eq.\,(1).
Finally, we can ignore the magnetic field dependence of $\pc$ and assume $\pc \equiv \pc^{(0)}\eac^2(\eac^2+1)/(\eac^2-1)^2\simeq \pc^{(0)}=e^2\ec^2 v_{F}^2/\omega^4$, where $\ec$ is the microwave field and $v_F$ is the Fermi velocity.

Using the data presented in Fig.\,\ref{fig1} we extract the normalized MIRO amplitude, $\delta \rho/\eac$, which, regardless of the model, is expected to scale with $\delta^2=\exp(-2\pi\eac/\omega\tauq)$.
The results for $T=1,\,2,\,3,\,4$ K are presented in Fig.\,2\,(a) as a function of $\eac$.
Having observed exponential dependences over at least two orders of magnitude in all data sets we make two important observations.
First, the slope, $-2\pi/\omega\tauq$, monotonically grows with $T$ by absolute value, marking the increase of the quantum scattering rate.
Second, all data sets can be fitted to converge to a single point at $\eac=0$, indicating that the pre-factor in Eq.\,(1) is essentially temperature independent [cf. inset of Fig.\,2\,(a)].

\begin{figure}[t]
\includegraphics{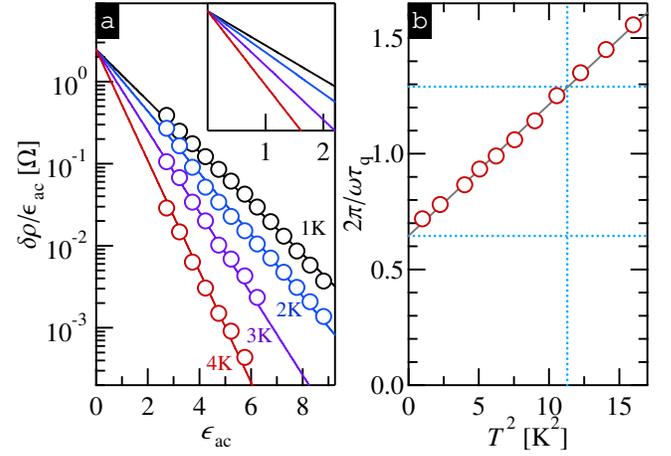}
\caption{(color online) 
(a) Normalized MIRO amplitude $\delta \rho/\eac$ vs. $\eac$ at $T =1.0,\,2.0,\,3.0,\,4.0$ K (circles) and fits to $\exp(-2\pi\eac/\omega\tauq)$ (lines). 
Inset shows that all fits intersect at $\eac=0$.
(b) Normalized quantum scattering rate $2\pi/\omega\tauq$ vs. $T^2$. 
Horizontal lines mark $\tauq=\tauim$ and $\tauq=\tauim/2$, satisfied at $T^2=0$ and $T^2\simeq 11$ K$^2$, respectively. 
}
\label{fig2}
\end{figure}
After repeating the Dingle plot procedure for other temperatures we present the extracted $2\pi/\omega\tauq$ vs. $T^2$ in Fig.\,\ref{fig2}\,(b).
Remarkably, the quantum scattering rate follows quadratic dependence over the whole range of temperatures studied.
This result is reminiscent of the temperature dependence of quantum lifetime in double quantum wells obtained by tunneling spectroscopy \citep{murphy:eisenstein} and from the analysis of the intersubband magnetoresistance oscillations \citep{berk:slutzky:mamani}.
In those experiments, it was suggested that the temperature dependence of $1/\tauq$ emerges from the electron-electron scattering, which is expected to greatly exceed the electron-phonon contribution. 
Here, we take the same approach and assume $1/\tauq=1/\tauim+1/\tauee$, where $\tauim$ and $\tauee$ are the impurity and electron-electron contributions, respectively.
Using the well-known estimate for the electron-electron scattering rate \citep{chaplik:giuliani}, $1/\tauee=\lambda T^2/\varepsilon_F$, where $\lambda$ is a constant of the order of unity, we perform the linear fit to the data in Fig.\,\ref{fig2}\,(b) and obtain $\tauim \simeq 19$ ps and $\lambda \simeq 4.1$.
We do not attempt a comparison of extracted $\tauim$ with the one obtained from SdHO analysis since the latter is known to severely underestimate this parameter. 

To confirm our conclusions we now plot in Fig.\,3\,(a) the normalized MIRO amplitude, $\delta \rho/\eac$, evaluated at the MIRO maxima near $\eac=n-1/4$ for $n=3,4,5,6$ as a function of $T^2$.
We observe that all data sets are well described by the exponential, $\exp(-\alpha T^2)$, over several orders of magnitude and that the exponent, $\alpha$, monotonically increases with $\eac$. 
The inset of Fig.\,3\,(a) shows the extension of the fits into the negative $T^2$ region revealing an intercept at $\simeq - 11$ K$^2$.
This intercept indicates that at $\bar T^2 \simeq 11 $ K$^2$, $\tauee \simeq \tauim$  providing an alternative way to estimate $\lambda$.
Indeed, direct examination of the data in Fig.\,2\,(b) reveals that the electron-electron contribution approaches the impurity contribution at $\bar T^2 \simeq 11 $ K$^2$, {\em i.e.} $1/\tauq(\bar T) = 1/\tauee(\bar T)+1/\tauim \simeq 2/\tauim=2/\tauq(0)$.
Another way to obtain parameter $\lambda$ is to extract the exponent, $\alpha$, from the data in Fig.\,3\,(a) and examine its dependence on $\eac$.
This is done in Fig.\,3\,(b) which shows the anticipated linear dependence, $\alpha=(2\pi\lambda/\omega\varepsilon_F)\eac$, from which we confirm $\lambda \simeq 4.1$.

\begin{figure}[t]
\includegraphics{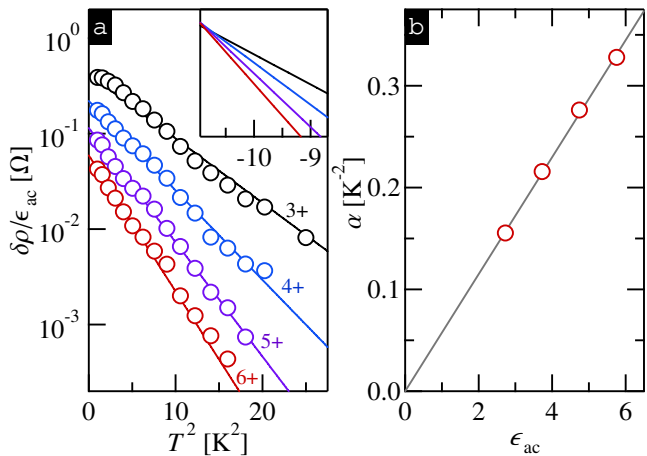}
\caption{(color online) 
(a) Normalized MIRO amplitude, $\delta \rho/\eac$, vs. $T^2$ near $\eac=2.75,\,3.75,\,4.75,\,5.75$ (circles) and fits to $\exp(-\alpha T^2)$ (lines). 
Inset demonstrates that all fits intersect at $-11$ K$^2$.
(b) Extracted exponent $\alpha$ vs. $\eac$ reveals expected linear dependence.
}
\label{fig3}
\end{figure}
To summarize our observations, the MIRO amplitude as a function of $T$ and $\eac$ is found to conform to a simple expression:
\begin{equation}
\delta \rho \simeq A \eac \exp [-2\pi/\oc\tauq].
\label{ampl}
\end{equation}
Here, $A$ is roughly independent on $T$, but $\tauq$ is temperature dependent due to electron-electron interactions:
\begin{equation}
\frac 1 \tauq = \frac 1 \tauim+\frac 1 \tauee, \,\,\, \frac 1 \tauee \simeq \lambda \frac {T^2}{\varepsilon_F}.
\label{tauq}
\end{equation}
It is illustrative to plot all our data as a function of $2\pi/\oc\tauq$, where $\tauq$ is evaluated using Eq.\,(\ref{tauq}).
As shown in Fig.\,4\,(a), when plotted in such a way, all the data collected at different temperatures collapse together to show universal exponential dependence over three orders of magnitude. 
The line in Fig.\,4\,(a), drawn with the slope of Eq.\,(\ref{tauq}), confirms excellent agreement over the whole range of $\eac$ and $T$.

\begin{figure}[t]
\includegraphics{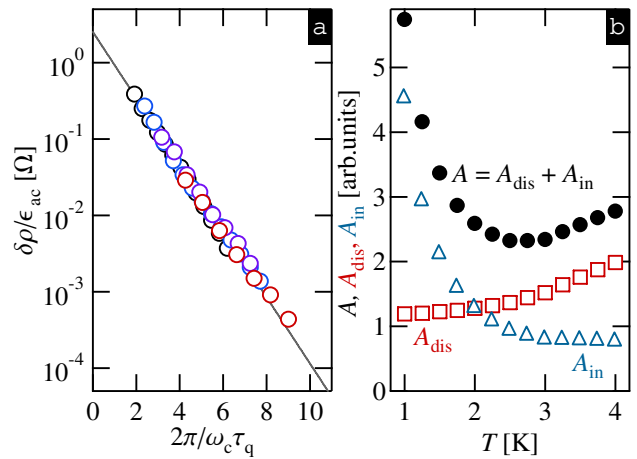}
\caption{(color online) 
(a) Normalized MIRO amplitude $\delta \rho/\eac$ vs. $2\pi/\omega\tauq$ for $T =1.0,\,2.0,\,3.0,\,4.0$ K (circles). 
Solid line marks a slope of $\exp(-2\pi/\oc\tauq)$.
(b) Contributions $\adis$ (squares), $\ain$ (triangles), and $A$ (circles) vs. $T$.
}
\label{fig4}
\end{figure}

We now discuss observed temperature independence of $A$, which we present as a sum of the ``displacement'' and the ``inelastic'' contributions, $A=\adis+\ain$.
According to Eq.\,(\ref{theory}), at low $T$ $\adis < \ain$ but at high $T$ $\adis > \ain$.
Therefore, there should exist a crossover temperature $T^*$, such that $\adis(T^*)=\ain(T^*)$. 
Assuming $\tauin \simeq \tauee \simeq \varepsilon_F/\lambda T^2 $ we obtain $T^*\simeq 2$ K and conclude that the ``displacement'' contribution cannot be ignored down to the lowest temperature studied.
Next, we notice that Eq.\,(\ref{theory}) contains transport scattering time, $\tautr$, which varies roughly by a factor of two in our temperature range.
If this variation is taken into account, $\ain$ will decay considerably slower than $1/T^2$ and $\adis$ will grow with $T$, instead of being $T$-independent, leading to a rather weak temperature dependence of $A$.
This is illustrated in Fig.\,\ref{fig4}(b) showing temperature evolution of both contributions and of their sum, which exhibits rather weak temperature dependence at $T\gtrsim 1.5$ K. 
In light of the temperature dependent exponent, we do not attempt to analyze this subtle behavior using our data.

Finally, we notice that the ``displacement'' contribution in Eq.\,(1) was obtained under the assumption of small-angle scattering caused by remote impurities.
However, it is known from non-linear transport measurements that short-range scatterers are intrinsic to high mobility structures \citep{yang:2002a,ac:dc}.
It is also established theoretically that including a small amount of short-range scatterers on top of the smooth background potential provides a better description of real high-mobility structures \citep{mirlin:gornyi}.
It is reasonable to expect that consideration of short-range scatterers will increase ``displacement'' contribution leading to lower $T^*$.

To summarize, we have studied MIRO temperature dependence in a high-mobility 2DES. 
We have found that the temperature dependence is exponential and originates from the temperature-dependent quantum lifetime entering the square of the Dingle factor.
The corresponding correction to the quantum scattering rate obeys $T^2$ dependence, consistent with the electron-electron interaction effects.
At the same time we are unable to identify any significant temperature dependence of the pre-factor in Eq.\,(1), which can be partially accounted for by the interplay between the ``displacement'' and the ``inelastic'' contributions in our high-mobility 2DES.
Since this observation might be unique to our structures, further systematic experiments in samples with different amounts and types of disorder are highly desirable.
It is also important to theoretically consider the effects of short-range impurity and electron-phonon scattering. 
Another important issue is the influence of the electron-electron interactions on single particle lifetime entering the square of the Dingle factor appearing in MIRO (which are different from the Shubnikov-de Haas oscillations where the Dingle factor does not contain the $1/\tauee \propto T^2$ term \citep{martin:adamov}).
We note that such a scenario was considered a few years ago \citep{ryzhii}.

We thank A. V. Chubukov, I. A. Dmitriev, R. R. Du, A. Kamenev, M. Khodas, A. D. Mirlin, F. von Oppen, D. G. Polyakov, M. E. Raikh, B. I. Shklovskii, and M. G. Vavilov for discussions and critical comments, and W. Zhang for contribution to initial experiments.
The work in Minnesota was supported by NSF Grant No. DMR-0548014.
\vspace{-0.15in}

\begin{thebibliography}{57}
\expandafter\ifx\csname natexlab\endcsname\relax\def\natexlab#1{#1}\fi
\expandafter\ifx\csname bibnamefont\endcsname\relax
  \def\bibnamefont#1{#1}\fi
\expandafter\ifx\csname bibfnamefont\endcsname\relax
  \def\bibfnamefont#1{#1}\fi
\expandafter\ifx\csname citenamefont\endcsname\relax
  \def\citenamefont#1{#1}\fi
\expandafter\ifx\csname url\endcsname\relax
  \def\url#1{\texttt{#1}}\fi
\expandafter\ifx\csname urlprefix\endcsname\relax\def\urlprefix{URL }\fi
\providecommand{\bibinfo}[2]{#2}
\providecommand{\eprint}[2][]{\url{#2}}

\bibitem[{\citenamefont{Shubnikov and de~Haas}(1930)}]{shubnikov:1930}
\bibinfo{author}{\bibfnamefont{L.}~\bibnamefont{Shubnikov}} \bibnamefont{and}
  \bibinfo{author}{\bibfnamefont{W.}~\bibnamefont{de~Haas}},
  \bibinfo{journal}{Leiden Commun.} \textbf{\bibinfo{volume}{207a}},
  \bibinfo{pages}{3} (\bibinfo{year}{1930}).

\bibitem[{mir({\natexlab{a}})}]{miro:exp}
 \bibinfo{author}{\bibfnamefont{M.~A.} \bibnamefont{Zudov}},
  \bibinfo{author}{\bibfnamefont{R.~R.} \bibnamefont{Du}},
  \bibinfo{author}{\bibfnamefont{J.~A.} \bibnamefont{Simmons}},
  \bibnamefont{and} \bibinfo{author}{\bibfnamefont{J.~L.} \bibnamefont{Reno}},
  \bibinfo{journal}{Phys. Rev. B} \textbf{\bibinfo{volume}{64}},
  \bibinfo{eid}{201311(R)} (\bibinfo{year}{2001}{\natexlab{a}});
\bibinfo{author}{\bibfnamefont{P.~D.} \bibnamefont{Ye}},
  \bibinfo{author}{\bibfnamefont{L.~W.} \bibnamefont{Engel}},
  \bibinfo{author}{\bibfnamefont{D.~C.} \bibnamefont{Tsui}},
  \bibinfo{author}{\bibfnamefont{J.~A.} \bibnamefont{Simmons}},
  \bibinfo{author}{\bibfnamefont{J.~R.} \bibnamefont{Wendt}},
  \bibinfo{author}{\bibfnamefont{G.~A.} \bibnamefont{Vawter}},
  \bibnamefont{and} \bibinfo{author}{\bibfnamefont{J.~L.} \bibnamefont{Reno}},
  \bibinfo{journal}{Appl. Phys. Lett.} \textbf{\bibinfo{volume}{79}},
  \bibinfo{eid}{2193} (\bibinfo{year}{2001}).  

\bibitem[{\citenamefont{Yang et~al.}(2002)\citenamefont{Yang, Zhang, Du,
  Simmons, and Reno}}]{yang:2002a}
\bibinfo{author}{\bibfnamefont{C.~L.} \bibnamefont{Yang}},
  \bibinfo{author}{\bibfnamefont{J.}~\bibnamefont{Zhang}},
  \bibinfo{author}{\bibfnamefont{R.~R.} \bibnamefont{Du}},
  \bibinfo{author}{\bibfnamefont{J.~A.} \bibnamefont{Simmons}},
  \bibnamefont{and} \bibinfo{author}{\bibfnamefont{J.~L.} \bibnamefont{Reno}},
  \bibinfo{journal}{Phys. Rev. Lett.} \textbf{\bibinfo{volume}{89}},
  \bibinfo{eid}{076801} (\bibinfo{year}{2002}).

\bibitem[{\citenamefont{Zudov et~al.}(2001{\natexlab{b}})\citenamefont{Zudov,
  Ponomarev, Efros, Du, Simmons, and Reno}}]{zudov:2001b}
\bibinfo{author}{\bibfnamefont{M.~A.} \bibnamefont{Zudov}},
  \bibinfo{author}{\bibfnamefont{I.~V.} \bibnamefont{Ponomarev}},
  \bibinfo{author}{\bibfnamefont{A.~L.} \bibnamefont{Efros}},
  \bibinfo{author}{\bibfnamefont{R.~R.} \bibnamefont{Du}},
  \bibinfo{author}{\bibfnamefont{J.~A.} \bibnamefont{Simmons}},
  \bibnamefont{and} \bibinfo{author}{\bibfnamefont{J.~L.} \bibnamefont{Reno}},
  \bibinfo{journal}{Phys. Rev. Lett.} \textbf{\bibinfo{volume}{86}},
  \bibinfo{pages}{3614} (\bibinfo{year}{2001}{\natexlab{b}}).

\bibitem[{\citenamefont{Mani et~al.}(2002)\citenamefont{Mani, Smet, von
  Klitzing, Narayanamurti, Johnson, and Umansky}}]{mani:2002}
\bibinfo{author}{\bibfnamefont{R.~G.} \bibnamefont{Mani}},
  \bibinfo{author}{\bibfnamefont{J.~H.} \bibnamefont{Smet}},
  \bibinfo{author}{\bibfnamefont{K.}~\bibnamefont{von Klitzing}},
  \bibinfo{author}{\bibfnamefont{V.}~\bibnamefont{Narayanamurti}},
  \bibinfo{author}{\bibfnamefont{W.~B.} \bibnamefont{Johnson}},
  \bibnamefont{and} \bibinfo{author}{\bibfnamefont{V.}~\bibnamefont{Umansky}},
  \bibinfo{journal}{Nature} \textbf{\bibinfo{volume}{420}},
  \bibinfo{pages}{646} (\bibinfo{year}{2002}).

\bibitem[{\citenamefont{Zudov et~al.}(2003)\citenamefont{Zudov, Du, Pfeiffer,
  and West}}]{zudov:2003}
\bibinfo{author}{\bibfnamefont{M.~A.} \bibnamefont{Zudov}},
  \bibinfo{author}{\bibfnamefont{R.~R.} \bibnamefont{Du}},
  \bibinfo{author}{\bibfnamefont{L.~N.} \bibnamefont{Pfeiffer}},
  \bibnamefont{and} \bibinfo{author}{\bibfnamefont{K.~W.} \bibnamefont{West}},
  \bibinfo{journal}{Phys. Rev. Lett.} \textbf{\bibinfo{volume}{90}},
  \bibinfo{eid}{046807} (\bibinfo{year}{2003}).

\bibitem[{\citenamefont{Willett et~al.}(2004)\citenamefont{Willett, Pfeiffer,
  and West}}]{willett:2004}
\bibinfo{author}{\bibfnamefont{R.~L.} \bibnamefont{Willett}},
  \bibinfo{author}{\bibfnamefont{L.~N.} \bibnamefont{Pfeiffer}},
  \bibnamefont{and} \bibinfo{author}{\bibfnamefont{K.~W.} \bibnamefont{West}},
  \bibinfo{journal}{Phys. Rev. Lett.} \textbf{\bibinfo{volume}{93}},
  \bibinfo{eid}{026804} (\bibinfo{year}{2004}).

\bibitem[{\citenamefont{Andreev et~al.}(2003)\citenamefont{Andreev, Aleiner,
  and Millis}}]{zrs:other}
\bibinfo{author}{\bibfnamefont{A.~V.} \bibnamefont{Andreev}},
  \bibinfo{author}{\bibfnamefont{I.~L.} \bibnamefont{Aleiner}},
  \bibnamefont{and} \bibinfo{author}{\bibfnamefont{A.~J.}
  \bibnamefont{Millis}}, \bibinfo{journal}{Phys. Rev. Lett.}
  \textbf{\bibinfo{volume}{91}}, \bibinfo{eid}{056803} (\bibinfo{year}{2003});
\bibinfo{author}{\bibfnamefont{C.~L.} \bibnamefont{Yang}},
  \bibinfo{author}{\bibfnamefont{M.~A.} \bibnamefont{Zudov}},
  \bibinfo{author}{\bibfnamefont{T.~A.} \bibnamefont{Knuuttila}},
  \bibinfo{author}{\bibfnamefont{R.~R.} \bibnamefont{Du}},
  \bibinfo{author}{\bibfnamefont{L.~N.} \bibnamefont{Pfeiffer}},
  \bibnamefont{and} \bibinfo{author}{\bibfnamefont{K.~W.} \bibnamefont{West}},
  \bibinfo{journal}{Phys. Rev. Lett.} \textbf{\bibinfo{volume}{91}},
  \bibinfo{eid}{096803} (\bibinfo{year}{2003});
\bibinfo{author}{\bibfnamefont{R.~G.} \bibnamefont{Mani}},
  \bibinfo{author}{\bibfnamefont{V.}~\bibnamefont{Narayanamurti}},
  \bibinfo{author}{\bibfnamefont{K.}~\bibnamefont{von Klitzing}},
  \bibinfo{author}{\bibfnamefont{J.~H.} \bibnamefont{Smet}},
  \bibinfo{author}{\bibfnamefont{W.~B.} \bibnamefont{Johnson}},
  \bibnamefont{and} \bibinfo{author}{\bibfnamefont{V.}~\bibnamefont{Umansky}},
  \bibinfo{journal}{Phys. Rev. B} \textbf{\bibinfo{volume}{70}},
  \bibinfo{eid}{155310} (\bibinfo{year}{2004}{\natexlab{a}});
\bibinfo{author}{\bibfnamefont{A.}~\bibnamefont{Auerbach}},
  \bibinfo{author}{\bibfnamefont{I.}~\bibnamefont{Finkler}},
  \bibinfo{author}{\bibfnamefont{B.~I.} \bibnamefont{Halperin}},
  \bibnamefont{and} \bibinfo{author}{\bibfnamefont{A.}~\bibnamefont{Yacoby}},
  \bibinfo{journal}{Phys. Rev. Lett.} \textbf{\bibinfo{volume}{94}},
  \bibinfo{eid}{196801} (\bibinfo{year}{2005});
\bibinfo{author}{\bibfnamefont{A.~A.} \bibnamefont{Bykov}},
  \bibinfo{author}{\bibfnamefont{A.~K.} \bibnamefont{Bakarov}},
  \bibinfo{author}{\bibfnamefont{D.~R.} \bibnamefont{Islamov}},
  \bibnamefont{and} \bibinfo{author}{\bibfnamefont{A.~I.}
  \bibnamefont{Toropov}}, \bibinfo{journal}{JETP Lett.}
  \textbf{\bibinfo{volume}{84}}, \bibinfo{pages}{391} (\bibinfo{year}{2006});
\bibinfo{author}{\bibfnamefont{M.~A.} \bibnamefont{Zudov}},
  \bibinfo{author}{\bibfnamefont{R.~R.} \bibnamefont{Du}},
  \bibinfo{author}{\bibfnamefont{L.~N.} \bibnamefont{Pfeiffer}},
  \bibnamefont{and} \bibinfo{author}{\bibfnamefont{K.~W.} \bibnamefont{West}},
  \bibinfo{journal}{Phys. Rev. B} \textbf{\bibinfo{volume}{73}},
  \bibinfo{eid}{041303(R)} (\bibinfo{year}{2006});
  \bibinfo{journal}{Phys. Rev. Lett.} \textbf{\bibinfo{volume}{96}},
  \bibinfo{eid}{236804} (\bibinfo{year}{2006}{\natexlab{b}}).

\bibitem[{\citenamefont{Bykov et~al.}(2007)\citenamefont{Bykov, Zhang,
  Vitkalov, Kalagin, and Bakarov}}]{bykov:zhang}
\bibinfo{author}{\bibfnamefont{A.~A.} \bibnamefont{Bykov}},
  \bibinfo{author}{\bibfnamefont{J.-Q.} \bibnamefont{Zhang}},
  \bibinfo{author}{\bibfnamefont{S.}~\bibnamefont{Vitkalov}},
  \bibinfo{author}{\bibfnamefont{A.~K.} \bibnamefont{Kalagin}},
  \bibnamefont{and} \bibinfo{author}{\bibfnamefont{A.~K.}
  \bibnamefont{Bakarov}}, \bibinfo{journal}{Phys. Rev. Lett.}
  \textbf{\bibinfo{volume}{99}}, \bibinfo{eid}{116801} (\bibinfo{year}{2007});
\bibinfo{author}{\bibfnamefont{W.}~\bibnamefont{Zhang}},
  \bibinfo{author}{\bibfnamefont{M.~A.} \bibnamefont{Zudov}},
  \bibinfo{author}{\bibfnamefont{L.~N.} \bibnamefont{Pfeiffer}},
  \bibnamefont{and} \bibinfo{author}{\bibfnamefont{K.~W.} \bibnamefont{West}},
  \bibinfo{journal}{Phys. Rev. Lett.} \textbf{\bibinfo{volume}{100}},
  \bibinfo{eid}{036805} (\bibinfo{year}{2008}).

\bibitem[{\citenamefont{Smet et~al.}(2005)\citenamefont{Smet, Gorshunov, Jiang,
  Pfeiffer, West, Umansky, Dressel, Meisels, Kuchar, and von
  Klitzing}}]{smet:2005}
\bibinfo{author}{\bibfnamefont{J.~H.} \bibnamefont{Smet}},
  \bibinfo{author}{\bibfnamefont{B.}~\bibnamefont{Gorshunov}},
  \bibinfo{author}{\bibfnamefont{C.}~\bibnamefont{Jiang}},
  \bibinfo{author}{\bibfnamefont{L.}~\bibnamefont{Pfeiffer}},
  \bibinfo{author}{\bibfnamefont{K.}~\bibnamefont{West}},
  \bibinfo{author}{\bibfnamefont{V.}~\bibnamefont{Umansky}},
  \bibinfo{author}{\bibfnamefont{M.}~\bibnamefont{Dressel}},
  \bibinfo{author}{\bibfnamefont{R.}~\bibnamefont{Meisels}},
  \bibinfo{author}{\bibfnamefont{F.}~\bibnamefont{Kuchar}}, \bibnamefont{and}
  \bibinfo{author}{\bibfnamefont{K.}~\bibnamefont{von Klitzing}},
  \bibinfo{journal}{Phys. Rev. Lett.} \textbf{\bibinfo{volume}{95}},
  \bibinfo{eid}{116804} (\bibinfo{year}{2005}).

\bibitem[{\citenamefont{Mani}(2005)}]{mani:yang}
\bibinfo{author}{\bibfnamefont{R.~G.} \bibnamefont{Mani}},
  \bibinfo{journal}{Phys. Rev. B} \textbf{\bibinfo{volume}{72}},
  \bibinfo{eid}{075327} (\bibinfo{year}{2005});
\bibinfo{author}{\bibfnamefont{C.~L.} \bibnamefont{Yang}},
  \bibinfo{author}{\bibfnamefont{R.~R.} \bibnamefont{Du}},
  \bibinfo{author}{\bibfnamefont{L.~N.} \bibnamefont{Pfeiffer}},
  \bibnamefont{and} \bibinfo{author}{\bibfnamefont{K.~W.} \bibnamefont{West}},
  \bibinfo{journal}{Phys. Rev. B} \textbf{\bibinfo{volume}{74}},
  \bibinfo{eid}{045315} (\bibinfo{year}{2006}).

\bibitem[{\citenamefont{Studenikin et~al.}(2007)\citenamefont{Studenikin,
  Sachrajda, Gupta, Wasilewski, Fedorych, Byszewski, Maude, Potemski, Hilke,
  West et~al.}}]{studenikin:2007}
\bibinfo{author}{\bibfnamefont{S.~A.} \bibnamefont{Studenikin}},
  \bibinfo{author}{\bibfnamefont{A.~S.} \bibnamefont{Sachrajda}},
  \bibinfo{author}{\bibfnamefont{J.~A.} \bibnamefont{Gupta}},
  \bibinfo{author}{\bibfnamefont{Z.~R.} \bibnamefont{Wasilewski}},
  \bibinfo{author}{\bibfnamefont{O.~M.} \bibnamefont{Fedorych}},
  \bibinfo{author}{\bibfnamefont{M.}~\bibnamefont{Byszewski}},
  \bibinfo{author}{\bibfnamefont{D.~K.} \bibnamefont{Maude}},
  \bibinfo{author}{\bibfnamefont{M.}~\bibnamefont{Potemski}},
  \bibinfo{author}{\bibfnamefont{M.}~\bibnamefont{Hilke}},
  \bibinfo{author}{\bibfnamefont{K.~W.} \bibnamefont{West}},
  \bibnamefont{et~al.}, \bibinfo{journal}{Phys. Rev. B}
  \textbf{\bibinfo{volume}{76}}, \bibinfo{eid}{165321} (\bibinfo{year}{2007}).

\bibitem[{\citenamefont{Zudov}(2004)}]{miro:phase}
\bibinfo{author}{\bibfnamefont{M.~A.} \bibnamefont{Zudov}},
  \bibinfo{journal}{Phys. Rev. B} \textbf{\bibinfo{volume}{69}},
  \bibinfo{eid}{041304(R)} (\bibinfo{year}{2004});
\bibinfo{author}{\bibfnamefont{R.~G.} \bibnamefont{Mani}},
  \bibinfo{author}{\bibfnamefont{J.~H.} \bibnamefont{Smet}},
  \bibinfo{author}{\bibfnamefont{K.}~\bibnamefont{von Klitzing}},
  \bibinfo{author}{\bibfnamefont{V.}~\bibnamefont{Narayanamurti}},
  \bibinfo{author}{\bibfnamefont{W.~B.} \bibnamefont{Johnson}},
  \bibnamefont{and} \bibinfo{author}{\bibfnamefont{V.}~\bibnamefont{Umansky}},
  \bibinfo{journal}{Phys. Rev. Lett.} \textbf{\bibinfo{volume}{92}},
  \bibinfo{eid}{146801} (\bibinfo{year}{2004}{\natexlab{b}});
\bibinfo{author}{\bibfnamefont{S.~A.} \bibnamefont{Studenikin}},
  \bibinfo{author}{\bibfnamefont{M.}~\bibnamefont{Potemski}},
  \bibinfo{author}{\bibfnamefont{A.}~\bibnamefont{Sachrajda}},
  \bibinfo{author}{\bibfnamefont{M.}~\bibnamefont{Hilke}},
  \bibinfo{author}{\bibfnamefont{L.~N.} \bibnamefont{Pfeiffer}},
  \bibnamefont{and} \bibinfo{author}{\bibfnamefont{K.~W.} \bibnamefont{West}},
  \bibinfo{journal}{Phys. Rev. B} \textbf{\bibinfo{volume}{71}},
  \bibinfo{eid}{245313} (\bibinfo{year}{2005}).

\bibitem[{\citenamefont{Ryzhii}(1970)}]{disp:th}
\bibinfo{author}{\bibfnamefont{V.~I.} \bibnamefont{Ryzhii}},
  \bibinfo{journal}{Sov. Phys. Solid State} \textbf{\bibinfo{volume}{11}},
  \bibinfo{pages}{2078} (\bibinfo{year}{1970});
\bibinfo{author}{\bibfnamefont{V.~I.} \bibnamefont{Ryzhii}},
  \bibinfo{author}{\bibfnamefont{R.~A.} \bibnamefont{Suris}}, \bibnamefont{and}
  \bibinfo{author}{\bibfnamefont{B.~S.} \bibnamefont{Shchamkhalova}},
  \bibinfo{journal}{Sov. Phys.} \textbf{\bibinfo{volume}{20}},
  \bibinfo{pages}{1299} (\bibinfo{year}{1986});
\bibinfo{author}{\bibfnamefont{A.~C.} \bibnamefont{Durst}},
  \bibinfo{author}{\bibfnamefont{S.}~\bibnamefont{Sachdev}},
  \bibinfo{author}{\bibfnamefont{N.}~\bibnamefont{Read}}, \bibnamefont{and}
  \bibinfo{author}{\bibfnamefont{S.~M.} \bibnamefont{Girvin}},
  \bibinfo{journal}{Phys. Rev. Lett.} \textbf{\bibinfo{volume}{91}},
  \bibinfo{eid}{086803} (\bibinfo{year}{2003});
\bibinfo{author}{\bibfnamefont{X.~L.} \bibnamefont{Lei}} \bibnamefont{and}
  \bibinfo{author}{\bibfnamefont{S.~Y.} \bibnamefont{Liu}},
  \bibinfo{journal}{Phys. Rev. Lett.} \textbf{\bibinfo{volume}{91}},
  \bibinfo{eid}{226805} (\bibinfo{year}{2003});
\bibinfo{author}{\bibfnamefont{J.}~\bibnamefont{Shi}} \bibnamefont{and}
  \bibinfo{author}{\bibfnamefont{X.~C.} \bibnamefont{Xie}},
  \bibinfo{journal}{Phys. Rev. Lett.} \textbf{\bibinfo{volume}{91}},
  \bibinfo{eid}{086801} (\bibinfo{year}{2003});
\bibinfo{author}{\bibfnamefont{M.~G.} \bibnamefont{Vavilov}} \bibnamefont{and}
  \bibinfo{author}{\bibfnamefont{I.~L.} \bibnamefont{Aleiner}},
  \bibinfo{journal}{Phys. Rev. B} \textbf{\bibinfo{volume}{69}},
  \bibinfo{eid}{035303} (\bibinfo{year}{2004}).

\bibitem[{\citenamefont{Dorozhkin}(2003)}]{dorozhkin:2003}
\bibinfo{author}{\bibfnamefont{S.~I.} \bibnamefont{Dorozhkin}},
  \bibinfo{journal}{JETP Lett.} \textbf{\bibinfo{volume}{77}},
  \bibinfo{pages}{577} (\bibinfo{year}{2003}).

\bibitem[{\citenamefont{Dmitriev et~al.}(2003)\citenamefont{Dmitriev, Mirlin,
  and Polyakov}}]{dmp}
\bibinfo{author}{\bibfnamefont{I.~A.} \bibnamefont{Dmitriev}},
  \bibinfo{author}{\bibfnamefont{A.~D.} \bibnamefont{Mirlin}},
  \bibnamefont{and} \bibinfo{author}{\bibfnamefont{D.~G.}
  \bibnamefont{Polyakov}}, \bibinfo{journal}{Phys. Rev. Lett.}
  \textbf{\bibinfo{volume}{91}}, \bibinfo{eid}{226802} (\bibinfo{year}{2003});
  \bibinfo{journal}{Phys. Rev. B}
  \textbf{\bibinfo{volume}{75}}, \bibinfo{eid}{245320}
  (\bibinfo{year}{2007});
  \bibinfo{journal}{Phys. Rev. Lett.}
  \textbf{\bibinfo{volume}{99}}, \bibinfo{eid}{206805}
  (\bibinfo{year}{2007}).

\bibitem[{\citenamefont{Dmitriev et~al.}(2005)\citenamefont{Dmitriev, Vavilov,
  Aleiner, Mirlin, and Polyakov}}]{dmitriev:2005}
\bibinfo{author}{\bibfnamefont{I.~A.} \bibnamefont{Dmitriev}},
  \bibinfo{author}{\bibfnamefont{M.~G.} \bibnamefont{Vavilov}},
  \bibinfo{author}{\bibfnamefont{I.~L.} \bibnamefont{Aleiner}},
  \bibinfo{author}{\bibfnamefont{A.~D.} \bibnamefont{Mirlin}},
  \bibnamefont{and} \bibinfo{author}{\bibfnamefont{D.~G.}
  \bibnamefont{Polyakov}}, \bibinfo{journal}{Phys. Rev. B}
  \textbf{\bibinfo{volume}{71}}, \bibinfo{eid}{115316} (\bibinfo{year}{2005}).

\bibitem[{\citenamefont{Stormer et~al.}(1990)\citenamefont{Stormer, Pfeiffer,
  Baldwin, and West}}]{stormer:mendez}
\bibinfo{author}{\bibfnamefont{H.~L.} \bibnamefont{Stormer}},
  \bibinfo{author}{\bibfnamefont{L.~N.} \bibnamefont{Pfeiffer}},
  \bibinfo{author}{\bibfnamefont{K.~W.} \bibnamefont{Baldwin}},
  \bibnamefont{and} \bibinfo{author}{\bibfnamefont{K.~W.} \bibnamefont{West}},
  \bibinfo{journal}{Phys. Rev. B} \textbf{\bibinfo{volume}{41}},
  \bibinfo{pages}{1278} (\bibinfo{year}{1990}); 
\bibinfo{author}{\bibfnamefont{E.~E.} \bibnamefont{Mendez}},
  \bibinfo{author}{\bibfnamefont{P.~J.} \bibnamefont{Price}}, \bibnamefont{and}
  \bibinfo{author}{\bibfnamefont{M.}~\bibnamefont{Heiblum}},
  \bibinfo{journal}{Appl. Phys. Lett.} \textbf{\bibinfo{volume}{45}},
  \bibinfo{pages}{294} (\bibinfo{year}{1984}).
  
\bibitem[{\citenamefont{Murphy et~al.}(1995)\citenamefont{Murphy, Eisenstein,
  Pfeiffer, and West}}]{murphy:eisenstein}
\bibinfo{author}{\bibfnamefont{S.~Q.} \bibnamefont{Murphy}},
  \bibinfo{author}{\bibfnamefont{J.~P.} \bibnamefont{Eisenstein}},
  \bibinfo{author}{\bibfnamefont{L.~N.} \bibnamefont{Pfeiffer}},
  \bibnamefont{and} \bibinfo{author}{\bibfnamefont{K.~W.} \bibnamefont{West}},
  \bibinfo{journal}{Phys. Rev. B} \textbf{\bibinfo{volume}{52}},
  \bibinfo{pages}{14825} (\bibinfo{year}{1995});
\bibinfo{author}{\bibfnamefont{J.~P.} \bibnamefont{Eisenstein}},
  \bibinfo{author}{\bibfnamefont{D.}~\bibnamefont{Syphers}},
  \bibinfo{author}{\bibfnamefont{L.~N.} \bibnamefont{Pfeiffer}},
  \bibnamefont{and} \bibinfo{author}{\bibfnamefont{K.~W.} \bibnamefont{West}},
  \bibinfo{journal}{S. S. Comm.} \textbf{\bibinfo{volume}{143}},
  \bibinfo{pages}{365} (\bibinfo{year}{2007}).

\bibitem[{\citenamefont{Berk et~al.}(1995)\citenamefont{Berk, Kamenev,
  Palevski, Pfeiffer, and West}}]{berk:slutzky:mamani}
\bibinfo{author}{\bibfnamefont{Y.}~\bibnamefont{Berk}},
  \bibinfo{author}{\bibfnamefont{A.}~\bibnamefont{Kamenev}},
  \bibinfo{author}{\bibfnamefont{A.}~\bibnamefont{Palevski}},
  \bibinfo{author}{\bibfnamefont{L.~N.} \bibnamefont{Pfeiffer}},
  \bibnamefont{and} \bibinfo{author}{\bibfnamefont{K.~W.} \bibnamefont{West}},
  \bibinfo{journal}{Phys. Rev. B} \textbf{\bibinfo{volume}{51}},
  \bibinfo{pages}{2604} (\bibinfo{year}{1995});
\bibinfo{author}{\bibfnamefont{M.}~\bibnamefont{Slutzky}},
  \bibinfo{author}{\bibfnamefont{O.}~\bibnamefont{Entin-Wohlman}},
  \bibinfo{author}{\bibfnamefont{Y.}~\bibnamefont{Berk}},
  \bibinfo{author}{\bibfnamefont{A.}~\bibnamefont{Palevski}}, \bibnamefont{and}
  \bibinfo{author}{\bibfnamefont{H.}~\bibnamefont{Shtrikman}},
  \bibinfo{journal}{Phys. Rev. B} \textbf{\bibinfo{volume}{53}},
  \bibinfo{pages}{4065} (\bibinfo{year}{1996});
\bibinfo{author}{\bibfnamefont{N.~C.} \bibnamefont{Mamani}},
  \bibinfo{author}{\bibfnamefont{G.~M.} \bibnamefont{Gusev}},
  \bibinfo{author}{\bibfnamefont{T.~E.} \bibnamefont{Lamas}},
  \bibinfo{author}{\bibfnamefont{A.~K.} \bibnamefont{Bakarov}},
  \bibnamefont{and} \bibinfo{author}{\bibfnamefont{O.~E.}
  \bibnamefont{Raichev}}, \bibinfo{journal}{Phys. Rev. B}
  \textbf{\bibinfo{volume}{77}}, \bibinfo{eid}{205327} (\bibinfo{year}{2008}).

\bibitem[{\citenamefont{Chaplik}(1971)}]{chaplik:giuliani}
\bibinfo{author}{\bibfnamefont{A.~V.} \bibnamefont{Chaplik}},
  \bibinfo{journal}{Sov. Phys. JETP} \textbf{\bibinfo{volume}{33}},
  \bibinfo{pages}{997} (\bibinfo{year}{1971});
\bibinfo{author}{\bibfnamefont{G.~F.} \bibnamefont{Giuliani}} \bibnamefont{and}
  \bibinfo{author}{\bibfnamefont{J.~J.} \bibnamefont{Quinn}},
  \bibinfo{journal}{Phys. Rev. B} \textbf{\bibinfo{volume}{26}},
  \bibinfo{pages}{4421} (\bibinfo{year}{1982}).

\bibitem[{\citenamefont{Zhang et~al.}(2007{\natexlab{a}})\citenamefont{Zhang,
  Chiang, Zudov, Pfeiffer, and West}}]{ac:dc}
\bibinfo{author}{\bibfnamefont{W.}~\bibnamefont{Zhang}},
  \bibinfo{author}{\bibfnamefont{H.-S.} \bibnamefont{Chiang}},
  \bibinfo{author}{\bibfnamefont{M.~A.} \bibnamefont{Zudov}},
  \bibinfo{author}{\bibfnamefont{L.~N.} \bibnamefont{Pfeiffer}},
  \bibnamefont{and} \bibinfo{author}{\bibfnamefont{K.~W.} \bibnamefont{West}},
  \bibinfo{journal}{Phys. Rev. B} \textbf{\bibinfo{volume}{75}},
  \bibinfo{eid}{041304(R)} (\bibinfo{year}{2007}{\natexlab{a}});
\bibinfo{author}{\bibfnamefont{W.}~\bibnamefont{Zhang}},
  \bibinfo{author}{\bibfnamefont{M.~A.} \bibnamefont{Zudov}},
  \bibinfo{author}{\bibfnamefont{L.~N.} \bibnamefont{Pfeiffer}},
  \bibnamefont{and} \bibinfo{author}{\bibfnamefont{K.~W.} \bibnamefont{West}},
  \bibinfo{journal}{Phys. Rev. Lett.} \textbf{\bibinfo{volume}{98}},
  \bibinfo{pages}{106804} (\bibinfo{year}{2007}{\natexlab{b}});
\bibinfo{author}{\bibfnamefont{A.~T.} \bibnamefont{Hatke}},
  \bibinfo{author}{\bibfnamefont{H.-S.} \bibnamefont{Chiang}},
  \bibinfo{author}{\bibfnamefont{M.~A.} \bibnamefont{Zudov}},
  \bibinfo{author}{\bibfnamefont{L.~N.} \bibnamefont{Pfeiffer}},
  \bibnamefont{and} \bibinfo{author}{\bibfnamefont{K.~W.} \bibnamefont{West}},
  \bibinfo{journal}{Phys. Rev. B} \textbf{\bibinfo{volume}{77}},
  \bibinfo{eid}{201304(R)} (\bibinfo{year}{2008});
  \bibinfo{journal}{Phys. Rev. Lett.} \textbf{\bibinfo{volume}{101}},
  \bibinfo{eid}{246811} (\bibinfo{year}{2008});  
\bibinfo{author}{\bibfnamefont{X.~L.} \bibnamefont{Lei}},
  \bibinfo{journal}{Appl. Phys. Lett.} \textbf{\bibinfo{volume}{90}},
  \bibinfo{pages}{132119} (\bibinfo{year}{2007}); arXiv:0810.2014v2;  
\bibinfo{author}{\bibfnamefont{M.~G.} \bibnamefont{Vavilov}},
  \bibinfo{author}{\bibfnamefont{I.~L.} \bibnamefont{Aleiner}},
  \bibnamefont{and} \bibinfo{author}{\bibfnamefont{L.~I.}
  \bibnamefont{Glazman}}, \bibinfo{journal}{Phys. Rev. B}
  \textbf{\bibinfo{volume}{76}}, \bibinfo{eid}{115331} (\bibinfo{year}{2007});
\bibinfo{author}{\bibfnamefont{X.~L.} \bibnamefont{Lei}} \bibnamefont{and}
  \bibinfo{author}{\bibfnamefont{S.~Y.} \bibnamefont{Liu}},
  \bibinfo{journal}{Appl. Phys. Lett.} \textbf{\bibinfo{volume}{93}},
  \bibinfo{eid}{082101} (\bibinfo{year}{2008});
\bibinfo{author}{\bibfnamefont{M.}~\bibnamefont{Torres}} \bibnamefont{and}
  \bibinfo{author}{\bibfnamefont{A.}~\bibnamefont{Kunold}},
  \bibinfo{journal}{J. of Phys. A} \textbf{\bibinfo{volume}{41}},
  \bibinfo{pages}{304036} (\bibinfo{year}{2008});
\bibinfo{author}{\bibfnamefont{M.}~\bibnamefont{Khodas}} \bibnamefont{and}
  \bibinfo{author}{\bibfnamefont{M.~G.} \bibnamefont{Vavilov}},
  \bibinfo{journal}{Phys. Rev. B} \textbf{\bibinfo{volume}{78}},
  \bibinfo{eid}{245319} (\bibinfo{year}{2008}).

\bibitem[{\citenamefont{Mirlin et~al.}(2001)\citenamefont{Mirlin, Polyakov,
  Evers, and Wolfle}}]{mirlin:gornyi}
\bibinfo{author}{\bibfnamefont{A.~D.} \bibnamefont{Mirlin}},
  \bibinfo{author}{\bibfnamefont{D.~G.} \bibnamefont{Polyakov}},
  \bibinfo{author}{\bibfnamefont{F.}~\bibnamefont{Evers}}, \bibnamefont{and}
  \bibinfo{author}{\bibfnamefont{P.}~\bibnamefont{Wolfle}},
  \bibinfo{journal}{Phys. Rev. Lett.} \textbf{\bibinfo{volume}{87}},
  \bibinfo{pages}{126805} (\bibinfo{year}{2001});
\bibinfo{author}{\bibfnamefont{I.~V.} \bibnamefont{Gornyi}} \bibnamefont{and}
  \bibinfo{author}{\bibfnamefont{A.~D.} \bibnamefont{Mirlin}},
  \bibinfo{journal}{Phys. Rev. B} \textbf{\bibinfo{volume}{69}},
  \bibinfo{pages}{045313} (\bibinfo{year}{2004}).

\bibitem[{\citenamefont{Martin et~al.}(2003)\citenamefont{Martin, Maslov, and
  Reizer}}]{martin:adamov}
\bibinfo{author}{\bibfnamefont{G.~W.} \bibnamefont{Martin}},
  \bibinfo{author}{\bibfnamefont{D.~L.} \bibnamefont{Maslov}},
  \bibnamefont{and} \bibinfo{author}{\bibfnamefont{M.~Y.}
  \bibnamefont{Reizer}}, \bibinfo{journal}{Phys. Rev. B}
  \textbf{\bibinfo{volume}{68}}, \bibinfo{eid}{241309(R)} (\bibinfo{year}{2003});
\bibinfo{author}{\bibfnamefont{Y.}~\bibnamefont{Adamov}},
  \bibinfo{author}{\bibfnamefont{I.~V.} \bibnamefont{Gornyi}},
  \bibnamefont{and} \bibinfo{author}{\bibfnamefont{A.~D.}
  \bibnamefont{Mirlin}}, \bibinfo{journal}{Phys. Rev. B}
  \textbf{\bibinfo{volume}{73}}, \bibinfo{eid}{045426} (\bibinfo{year}{2006}).

\bibitem[{\citenamefont{Ryzhii and Suris}(2003)}]{ryzhii}
\bibinfo{author}{\bibfnamefont{V.}~\bibnamefont{Ryzhii}} \bibnamefont{and}
  \bibinfo{author}{\bibfnamefont{R.}~\bibnamefont{Suris}}, \bibinfo{journal}{J.
  Phys.: Condens. Matter} \textbf{\bibinfo{volume}{15}}, \bibinfo{pages}{6855}
  (\bibinfo{year}{2003});
\bibinfo{author}{\bibfnamefont{V.}~\bibnamefont{Ryzhii}},
  \bibinfo{author}{\bibfnamefont{A.}~\bibnamefont{Chaplik}}, \bibnamefont{and}
  \bibinfo{author}{\bibfnamefont{R.}~\bibnamefont{Suris}},
  \bibinfo{journal}{JETP Lett.} \textbf{\bibinfo{volume}{80}},
  \bibinfo{pages}{363} (\bibinfo{year}{2004}).

\end{thebibliography}

\end{document}